\documentclass[twocolumn,showpacs,preprintnumbers,nofootinbib,nobibnotes,amsmath,amssymb,aps,prl]{revtex4-1}
\usepackage{todonotes}
\usepackage{graphicx}
\usepackage{hyperref}
\usepackage{slashed}
\usepackage{amssymb,graphicx}
\usepackage{epstopdf}
\usepackage{amsmath,amsfonts}
\usepackage{epsfig} 
\usepackage{graphicx,graphics}

\usepackage{xcolor}

\begin{document} 

\begin{flushright}
  INR-TH-2021-009
\end{flushright}

\sloppy

\title{\bf Beyond freeze-in: Dark Matter via\\ inverse phase transition and gravitational wave signal}
\author{S.~Ramazanov$^{a}$, E.~Babichev$^{b}$, D.~Gorbunov$^{c,d}$, A.~Vikman$^{a}$\\
\small{\em $^a$CEICO, FZU-Institute of Physics of the Czech Academy of Sciences,}\\
\small{\em Na Slovance 1999/2, 182 21 Prague 8, Czech Republic}\\
\small{\em $^b$Universit\'e Paris-Saclay, CNRS/IN2P3, IJCLab, 91405 Orsay, France}\\
\small{\em $^c$Institute for Nuclear Research of the Russian Academy of Sciences, Moscow 117312, Russia}\\ 
\small{\em $^d$Moscow Institute of Physics and Technology, Dolgoprudny 141700, Russia}
}

\begin{abstract}

We propose a novel scenario of Dark Matter production naturally connected with generation of gravitational waves. Dark Matter is modelled as a real scalar, which interacts with the hot primordial plasma through a portal coupling to another scalar field. For a particular sign of the coupling, this system exhibits an inverse second order phase transition. The latter leads to an abundant Dark Matter production, even if the portal interaction is so weak that the freeze-in mechanism is inefficient. 
The model predicts domain wall formation in the Universe, long time before the inverse phase transition. These domain walls have a tension decreasing with time, and completely disappear at the inverse phase transition, so that the problem of overclosing the Universe is avoided. The domain wall network emits gravitational waves with characteristics defined by those of Dark Matter. In particular, the peak frequency of gravitational waves is determined by the portal coupling constant, and falls in the observable range for currently planned gravitational wave detectors.

\end{abstract}

{\let\newpage\relax\maketitle}

{\it Introduction and summary.} Freeze-out and freeze-in mechanisms are the most common ways of generating Dark Matter (DM) in the early Universe~\cite{Arcadi:2017kky, Bernal:2017kxu}. 
The former operates in the situation, when DM particles are in thermal equilibrium with the primordial plasma and requires 
relatively large couplings to (beyond) the Standard Model (SM)
species. For smaller coupling constants, the 
equilibrium may not be attained, but the observed DM abundance can be saturated via the freeze-in mechanism~\cite{McDonald:2001vt, Hall:2009bx}, i.e., by the out-of-equilibrium scatterings and decays of other particles. In the present work, we consider even weaker interactions than those assumed in freeze-in scenarios. 
Remarkably, not only efficient DM production is possible in that case, but also the feeble couplings involved can be probed in currently planned gravitational wave (GW) observations.

 We demonstrate this in the scenario of the scalar portal DM. Namely,
 it is assumed that the DM comprising a real scalar $\chi$ interacts
 with another scalar field (or a multiplet of fields) $\phi$, e.g.,
 the Higgs field, through the term $g^2 \chi^2 \phi^{\dagger}
 \phi/2$. The strength of the interaction is quantified by the
 dimensionless constant $g$. Stability of DM is protected by
 $Z_2$-symmetry. The field $\phi$ is assumed to be in equilibrium with
 (beyond) the SM degrees of freedom. For relatively large and
   moderate 
 constants, $g \gtrsim 10^{-5}-10^{-6}$, the dominant contribution to DM is yielded by freeze-out or freeze-in mechanisms~\cite{Chu:2011be, Yaguna:2011qn, Lebedev:2019ton}. 
 They fail to generate a right amount of DM for much weaker coupling, $g \ll 10^{-6}$. In this Letter we show how to overcome this difficulty: 
 the right abundance of DM can be produced even for exponentially smaller $g$ through an inverse phase transition. 
 The latter is a generic phenomenon which has been already discussed as early as in Ref.~\cite{Weinberg:1974hy}.
Furthermore, the inverse phase transition was used in Refs.~\cite{Dodelson:1989ii, Dodelson:1989cq} to tackle the problem of baryon asymmetry of the Universe. 
In the present work, we investigate DM applications of the inverse
phase transition continuing the line of research initiated in
Refs.~\cite{Babichev:2020xeg, Ramazanov:2020ajq}. While those
  works focus on DM interactions with gravity~\cite{Babichev:2020xeg} and primordial magnetic fields~\cite{Ramazanov:2020ajq}, here we discuss, how the inverse phase transition occurs in a more traditional setup of the portal DM interaction.

The mechanism of DM production, we propose in this Letter, is the following. Thermal fluctuations of the field $\phi$ described by 
variance $\sqrt{\langle \phi^{\dagger} \phi \rangle_{T}} \sim T$ give a contribution $\sim g^2 T^2$ to the DM effective mass squared on top of the bare mass squared $M^2$. 
For a particular sign in front of $g^2$-term this thermal contribution is negative. Hence, for sufficiently high temperatures spontaneous breaking of $Z_2$-symmetry
occurs, and the DM field $\chi$ takes a non-zero expectation value at a minimum of the effective potential. 
Later on, the thermal contribution to the total mass decreases enough
for the symmetry restoration (hence, inverse phase transition). Slightly before this moment, $\chi$ departures from the minimum and starts oscillating, which introduces the 
DM component in the Universe.

An interesting consequence of this DM genesis scenario is the formation of topological defects in the early Universe. Indeed, if the $Z_2$-symmetry was still unbroken, $\chi=0$, at the very origin of the hot big bang, then domain walls are inevitably formed, as the Universe reheats (and $Z_2$-symmetry breaks spontaneously). Our model provides a built-in mechanism for a fast thawing of domain walls: their tension $\sigma_{wall} (T)$ is determined
by thermal fluctuations of particles $\phi$, i.e., $\sigma_{wall} (T) \propto \langle \phi^{\dagger} \phi \rangle^{3/2}_{T}$, 
and thus drops with time as $\sigma_{wall} (T) \propto T^3$, see Refs.~\cite{Vilenkin:1981zs, Vilenkin}. As a result, the energy density of domain 
walls always degrades faster than radiation. They completely disappear at the inverse phase transition (cf., Ref.~\cite{Bettoni:2019dcw}). This is in contrast to the standard scenario of domain walls with a constant tension, which overclose the Universe, unless one introduces an explicit breaking of $Z_2$-symmetry~\cite{Zeldovich:1974uw, Gelmini:1988sf, Krajewski:2021jje} or creates a population bias of one degenerate vacuum over another~\cite{Coulson:1995nv}.

\begin{figure}[tb!]
  \begin{center}
    \includegraphics[width=\columnwidth,angle=0]{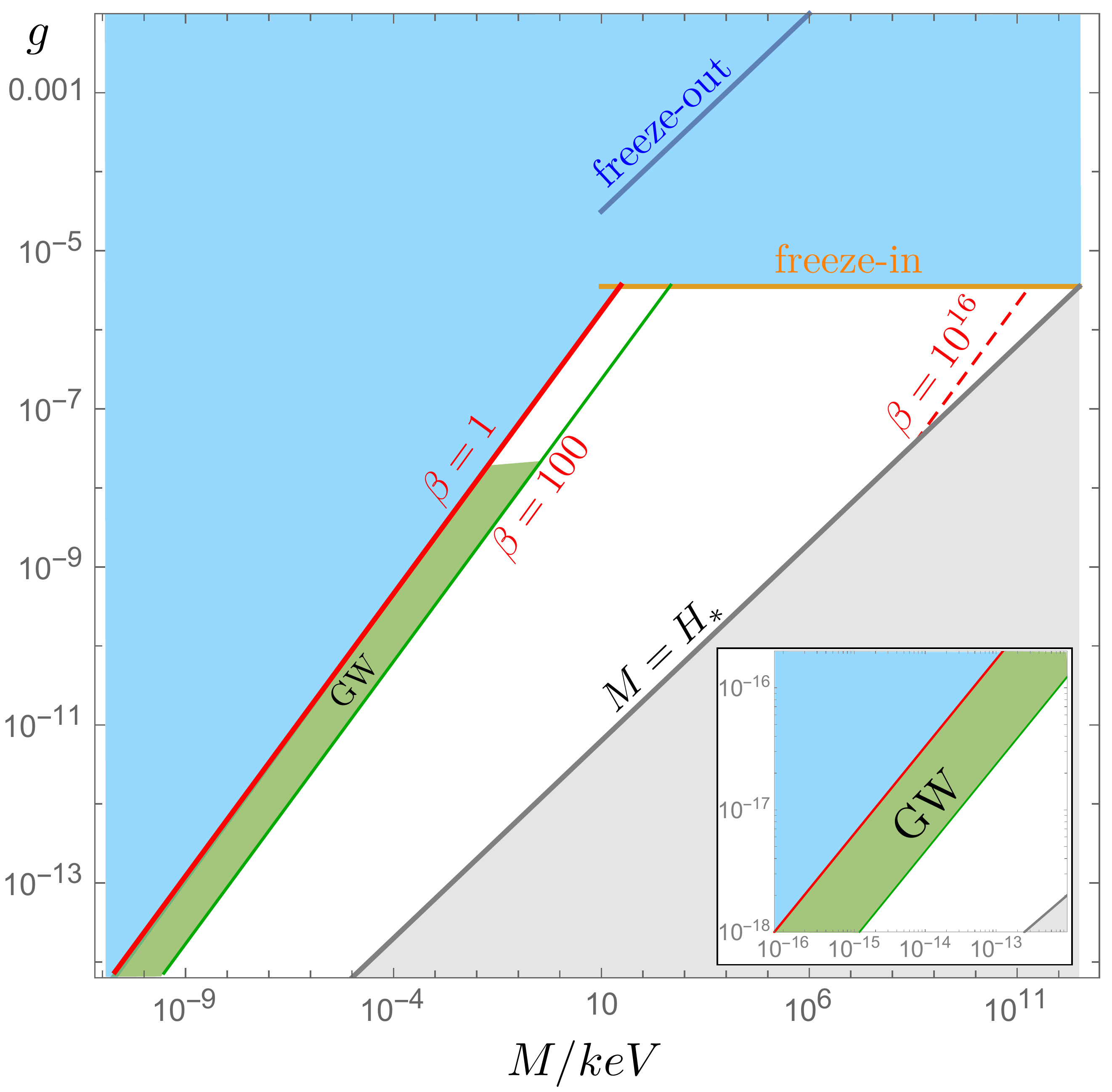}
  \caption{The region of space ($M$,\,$g$), where the observed DM
    abundance can be produced via the inverse phase transition is shown in
    white and green. We set $N=4$ and assumed that $\beta$ satisfies inequality \eqref{perturbativity}. Parameter space accessible by planned GW detectors is marked with green. For comparison, freeze-out and freeze-in 
  DM genesis scenarios are depicted with solid blue and orange lines,
  respectively. To plot them, we used expressions in
  Refs.~\cite{Chu:2011be} and~\cite{Lebedev:2019ton} and took masslesss
  $\phi$. The blue region is disfavored, because it leads either to over-/underproduction of DM composed of particles $\chi$ or runaway solutions in the $(\chi, \phi)$ field space. Our analysis of DM production via the inverse 
  phase transition relies on the condition $M>H_*$, it is inapplicable
  in the gray region where the phase transition does not occur.}
  \label{parameters}
  \end{center}
\end{figure}

\begin{figure}[tb!]
  \begin{center}
    \includegraphics[width=\columnwidth,angle=0]{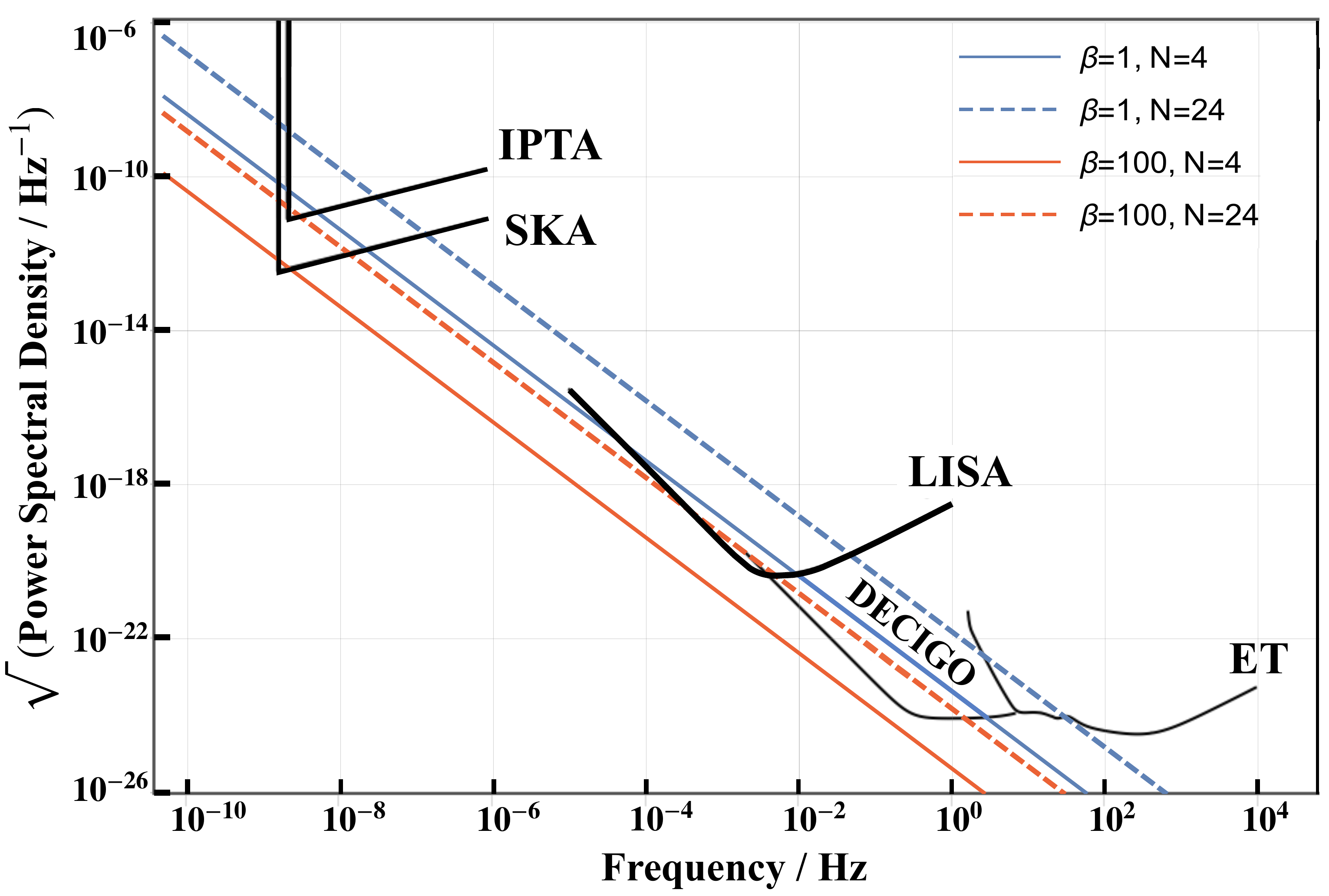}
  \caption{The root power spectral density $\sqrt{S_{h}}$ (strain) of GWs emitted by the network of domain walls is shown with straight colored lines as a function of the peak frequency $f_{gw}$ for different values of $\beta$ and $N$. 
Note that each particular set of parameters $(\beta, g)$ corresponds to one point on a coloured line, indicating the peak contribution to GWs given by Eq.~\eqref{today} for these parameters. Sensitivities of various GW detectors (black curved lines) have been plotted using the online tool~\cite{gwplotter}. }\label{waves}
  \end{center}
\end{figure}

The existence of domain walls in our scenario is interesting from the phenomenological point of view. The domain wall network in the early Universe serves as a source of stochastic GWs~\cite{Hiramatsu:2013qaa, Gleiser:1998na}. 
The present day peak frequency of GWs is estimated as $f_{gw} \simeq 100~\mbox{Hz} \cdot (g/10^{-8})$. Remarkably, the smaller the constant $g$ is, the better observational properties of the model are with respect to GWs.
In particular, values $g \lesssim 10^{-8}$, being characteristic for DM production 
through the inverse phase transition, see Fig.~\ref{parameters}, correspond to the frequencies $f_{gw} \lesssim 100~\mbox{Hz}$. For these frequencies, the model is testable with the future GW detectors, see Fig.~\ref{waves}.

{\it Dark Matter via inverse phase transition.} We consider the Lagrangian describing the DM field $\chi$, which we assume to be a real scalar singlet: 
\begin{equation}
\label{basic}
{\cal L}_{\chi}= \frac{(\partial {\chi})^2}{2}-\frac{M^2 \chi^2 }{2} -\frac{\lambda \chi^4 }{4} +\frac{g^2 \chi^2 \phi^{\dagger} \phi}{2} \; .
\end{equation}
Here $M$ and $\lambda$ denote the mass and the self-interaction coupling constant of the DM field;  $g^2$ is the portal coupling constant; $\phi$ is a scalar (or a scalar multiplet) being in thermal equilibrium with the primordial plasma. In the present work we crucially assume that the coupling constant $g^2$ between the fields $\chi$ and $\phi$ is positive: 
\begin{equation}
g^2 >0 \; .
\end{equation}
Note that the sign of $g^2$ is irrelevant in the freeze-out(in) mechanisms
of DM genesis, but it plays a key role in our case with the inverse phase transition\footnote{Flipping the signs of both $g^2$ and $M^2$ leads to a direct second order phase transition followed by formation of \emph{constant tension} domain walls that overclose the Universe, see, e.g., Ref.~\cite{Kolb:1990vq}.}. 
The above choice of sign for $g^2$ implies negative energy contribution from the interaction term. 
To avoid the runaway solutions in the space $(\chi, \phi)$, the following constraint must be obeyed: 
\begin{equation}
  \label{perturbativity}
\beta \equiv \frac{\lambda}{g^4}\geq \frac{1}{\lambda_\phi}\gtrsim 1 \; ,
\end{equation}
where $\lambda_{\phi}$ is the quartic self-interaction coupling of
$\phi$, and a weak coupling regime is assumed.

Provided that particles $\phi$ are relativistic at the relevant times, one has~\cite{Mukhanov}
\begin{equation}
\label{thermalaverage}
\langle \phi^{\dagger} \phi \rangle_{T} \approx \frac{NT^2}{12} \; ,
\end{equation}
where $N$ is the number of degrees of freedom associated with the field $\phi$. We write the resulting effective potential of the field $\chi$ as follows: 
\begin{equation}
\label{spontaneous}
V_{eff}=\frac{M^2\chi^2}{2}+\frac{\lambda \cdot \left(\chi^2-\eta^2 (T) \right)^2 }{4} \; ,
\end{equation}
where 
\begin{equation}
\label{tension}
\eta^2 (T) \approx \frac{Ng^2 T^2}{12 \lambda}  \; ,
\end{equation}
and we ignore the irrelevant difference between Eqs.~\eqref{basic}
and~(\ref{spontaneous}) due to the $\chi$-independent term. As it follows from Eqs.~\eqref{perturbativity} and~\eqref{tension}, backreaction of the field $\chi$ on the dynamics of the field $\phi$ is negligible. Indeed, the thermal mass squared of the field $\phi$ is constrained as $m^2_{\phi} \gtrsim N\lambda_{\phi} T^2/12$. It always exceeds the contribution $g^2 \chi^2 \approx NT^2/(12 \beta)$ following from the interaction with the DM field, provided the stability constraint~\eqref{perturbativity} is obeyed.

 When the temperature of the Universe 
is sufficiently high, $Z_2$-symmetry of the model is spontaneously broken, as it is clear from Eq.~\eqref{spontaneous}. In the spontaneously broken phase, the expectation value of $\chi$ is given by the minimum of the effective potential
\begin{equation}
\label{vacuum}
\langle \chi \rangle=\pm \sqrt{\eta^2 (T)-\frac{M^2}{\lambda}} \; .
\end{equation}
At the time $t=t_{sym}$, when the thermal mass becomes equal to the
bare mass of the field $\chi$, 
\begin{equation}
\label{transitionpoint}
M^2 =\lambda \eta^2 (T_{sym}) \approx \frac{Ng^2 T^2_{sym}}{12} \; ,
\end{equation}
the symmetry gets restored. Here we deal 
with the inverse second order phase transition\footnote{For DM production during the first order phase transition see Ref.~\cite{Azatov:2021ifm}.}, because the vacuum expectation value $\langle \chi \rangle$ continuously approaches zero, while its derivative explodes at $t=t_{sym}$. This fact plays a key role in what follows. 
Were the field $\chi$ tracking the minimum of the effective potential, Eq.~\eqref{vacuum}, down to $\chi=0$, its time derivative would also explode at $t=t_{sym}$. 
In practice, however, the field $\chi$ is kicked out of the minimum at some time $t_*$ slightly before $t_{sym}$, i.e., 
$t_{*} \lesssim t_{sym}$ (still, $t_* \approx t_{sym}$). Provided that $H_* \lesssim M$, where $H_*$ is the Hubble parameter at $t_*$, the field $\chi$ starts oscillating with the initial amplitude $\chi_*$ 
estimated as the vacuum expectation value at the time $t_*$, i.e., $\chi_* \simeq \langle \chi_* \rangle$. The expectation value $\langle \chi_* \rangle$ is inferred from the behaviour of the effective mass squared of the field $\chi$. In the broken phase, where $\chi \approx \langle \chi \rangle$, it is given by 
\begin{equation}
M^2_{eff}=3\lambda \chi^2-\lambda \eta^2 (T) +M^2 \approx 2\lambda \langle \chi \rangle^2 \; .
\end{equation}
At very early times, the effective mass changes adiabatically with time,
\begin{equation}
\label{ineq}
\left| \frac{\dot{M}_{eff}}{M^2_{eff}} \right| \ll 1\; ,
\end{equation}
and the field $\chi$ simply tracks its minimum, $\chi \approx
\langle \chi \rangle$. The tracking regime terminates at the time
$t_*$, when the condition~\eqref{ineq} becomes the equality. At this
moment  
the field $\chi$ value is estimated as
\begin{equation}
\label{init}
\chi_* \simeq \frac{\left(2M^{2}\right)^{1/3}}{\sqrt{2\lambda}}\left|\frac{\dot{\eta}}{\eta}\right|_*^{1/3}=\frac{(2M^2 H_*)^{1/3}}{\sqrt{2\lambda}} \; .
\end{equation}
The amplitude $\chi_*$ defines the DM energy density as
\begin{equation}
\label{energy}
\rho_{\chi} (t) \simeq \frac{M^2\chi^2_*}{2}  \cdot \left(\frac{a_*}{a (t)} \right)^3 \; .
\end{equation}
Remarkably, the estimate~\eqref{init}, checked numerically in Ref.~\cite{Babichev:2020xeg}, is largely model-independent. 

We are interested in the situation, when most of DM in the Universe is produced through the inverse phase transition\,\eqref{perturbativity}. This gives the constraint on the parameter space:
\begin{equation}
\label{constr2}
M \simeq 15~\mbox{eV} \cdot \frac{\beta^{3/5}}{\sqrt{N}} \cdot \left(\frac{g_* (T_*)}{100} \right)^{2/5} \cdot \left(\frac{g}{10^{-8}} \right)^{7/5} \; ,
\end{equation}
where $g_* (T)$ is the number of ultra-relativistic degrees of freedom in the Universe at the temperature $T$. The parameters $g$ and $M$ leading to the right abundance of DM through the inverse phase transition are shown in Fig.~\ref{parameters} for different values of $\beta$. The lines corresponding to the freeze-out and freeze-in mechanism are shown there for comparison. We see that the inverse phase transition largely extends the range of values of the portal coupling constant $g^2$, for which the right DM 
abundance can be obtained. Furthermore, our mechanism works in a wide range of values of the self-interaction constant $\lambda$ varying from $\lambda \sim g^4$ to $\lambda \sim 10^{-4}$ (corresponding to the right upper corner 
of the white space in Fig.~1).

{\it Formation of domain walls.} In the situation of our primary interest, when the DM field is feebly coupled to other matter fields, commonly discussed direct and indirect 
methods of catching DM particles may not work. This may raise doubts, if the case of very small $g $ is testable at all. However, if the field $\chi$ is set to zero, $\chi=0$, at the end of inflation, our mechanism of DM production 
leads to formation of domain walls in the early Universe. Indeed, there is no preference between positive and negative vacuum values, so that the field $\chi$ picks random values in different Hubble patches. Regions with 
different vacuum values are separated by domain walls. These domain walls generate GWs potentially observable with the near future experimental facilities.

The domain walls form only after the thermal DM mass $\sqrt{\lambda}\eta$ becomes comparable to the Hubble parameter. 
At earlier times the field $\chi$ is pinned to zero by the Hubble friction. Assuming that rolling to the minimum is substantially fast, i.e., occurs within a few Hubble times, we estimate the temperature at domain walls formation: 
\begin{equation}
\label{initial}
T_i \simeq  \frac{ \sqrt{N} gM_{Pl}}{\sqrt{g_* (T_i)}} \; .
\end{equation}
We use the reduced Planck mass $M_{Pl} \approx 2.44 \cdot 10^{18}~\mbox{GeV}$. Note that a finite duration of the roll results into the reduction of $T_i$ by the factor $\ln \left(2\eta_h/\delta \chi_h \right)$ compared to Eq.~\eqref{initial}, 
where $\eta_h$ and $\delta \chi_h$ are the expectation value and the perturbation of the field $\chi$ at the onset of the roll, time $t_h$. The perturbation $\delta \chi_h$ is determined by the past history of the field $\chi$ at inflation and 
preheating. In this work, we assume that $\delta \chi_h$ is not dramatically smaller than $\eta_h$, so that the logarithmic suppression can be safely ignored.

The tension of domain walls is given by\footnote{See, e.g., Ref.~\cite{Kolb:1990vq}, Eq.~(7.38) on page 215.} 
\begin{equation}
\label{surface}
\sigma_{wall} = \frac{2\sqrt{2\lambda} \eta^3 (T)}{3} \; .
\end{equation}
In the scaling regime~\cite{Press:1989yh, Garagounis:2002kt,
  Leite:2012vn}, there is one (or a few) domain wall(s) per horizon
volume, and the domain wall mass inside the Hubble radius reads
$M_{wall} \sim \sigma_{wall}/H^2$. The energy density of domain walls is estimated as $\rho_{wall} \sim M_{wall} H^3 \sim \sigma_{wall} H$. Using Eqs.~\eqref{tension},~\eqref{initial}, and~\eqref{surface}, we obtain the fraction of domain walls in the total energy budget of the Universe during radiation domination:
\begin{equation}
\frac{\rho_{wall}}{\rho_{rad}} \sim \frac{N^2}{30 g_{*} (T) \beta} \cdot \frac{T}{T_i} <1\; .
\end{equation}
The above inequality is always satisfied for not overly extensive $N$. Hence, the domain walls problem is 
automatically avoided in our setup.

{\it Properties of GWs.} Let us consider the GWs emitted by the network of thawing domain walls. We closely follow the discussion of Ref.~\cite{Hiramatsu:2013qaa}, which analyses properties of 
GWs by running lattice simulations in a scenario with constant
tension domain walls. 
We assume, that this analysis qualitatively captures features of GW production also in our case of a time-varying 
tension. This assumption is strongly supported by the fact that in both setups most energetic GWs are emitted in a short time interval: 
just before the domain wall collapse in Ref.~\cite{Hiramatsu:2013qaa} and right after domain wall formation in our case (see below). 
Approximating the tension to be constant in this short time interval, one validates the application of the results of Ref.~~\cite{Hiramatsu:2013qaa} to our case.
  
According to the Einstein quadrupole formula, the power of
gravitational radiation emission is estimated as $P \sim
\dddot{Q}^2_{ij}/(40\pi M^2_{Pl})$. In the scaling regime, the created
quadrupole moment is related to the wall mass $M_{wall}$
inside the Hubble radius $1/H$ by $|Q_{ij}|  \sim M_{wall}/
H^2$. Using $M_{wall}\sim \sigma_{wall} /H^2$, we obtain the estimate of the energy density of gravitational waves emitted at the time t:  
\begin{equation}
\rho_{gw} \sim P \cdot t \cdot H^3 \sim \frac{\sigma^2_{wall}}{40\pi M^2_{Pl}} \; ,
\end{equation}
so that $\rho_{gw} \propto T^6$ in our case.
We observe that GWs are most efficiently produced at high
temperatures $T \simeq T_i$
close to the moment of domain wall formation, provided that the scaling regime
is attained almost instantly. In the latter approximation, an accurate fit
to numerical simulations of Ref.~\cite{Hiramatsu:2013qaa} reads
\begin{equation}
\label{fractiongen}
\Omega_{gw} (t_i) \approx \frac{\lambda \tilde{\epsilon}_{gw} {\cal A}^2 \eta^6_i}{27\pi H^2_i M^4_{Pl}} \; ,
\end{equation}
where $\Omega_{gw} (t)$ is the fractional energy density of GWs per logarithmic frequency at peak. The factors $\tilde{\epsilon}_{gw}$ and ${\cal A}$ measure efficiency of GWs emission and the scaling property, correspondingly. 
Both turn out to be quite close to unity: $\tilde{\epsilon}_{gw}=0.7 \pm 0.4$ and ${\cal A}=0.8 \pm 0.1$. We proceed with central values of $\tilde{\epsilon}_{gw}$ and ${\cal A}$. 
Substituting \eqref{tension} into \eqref{fractiongen} and using \eqref{initial}, 
we get 
\begin{equation}
\label{atproduction}
\Omega_{gw} (t_i) \approx \frac{3 \cdot 10^{-9}\cdot N^4}{\beta^2} \cdot \left(\frac{100}{g_* (T_i)} \right)^2 \; .
\end{equation} 
During matter domination it redshifts as $\Omega_{gw} \sim 1/a$, so that presently we have
\begin{equation}
\label{today}
\Omega_{gw} h^2 (t_0) \approx \frac{4\cdot 10^{-14} \cdot N^4}{\beta^2} \cdot \left(\frac{100}{g_* (T_i)} \right)^{7/3} \, .
\end{equation}
The GW spectrum is peaked at frequency $f_{gw} (t_i) \simeq H_i$~\cite{Hiramatsu:2013qaa}, where $H_i$ is the Hubble rate at domain wall formation. Given  
Eq.~\eqref{initial}, the present day frequency of GWs reads
\begin{equation}
\label{frequency}
f_{gw}\equiv f_{gw} (t_0) \simeq 60~\mbox{Hz} \cdot \sqrt{N} \cdot \frac{g}{10^{-8}} \cdot \left(\frac{100}{g_* (T_i)} \right)^{1/3} \; .
\end{equation}
It is solely determined by the constant $g$. Equations
  \eqref{today}, \eqref{frequency} can be used to measure (bound) $g$
  and $\beta/N^2$ from positive (negative) results of GW searches at future detectors.

{\it Probing DM couplings with GWs.} Hereafter we restrict the
discussion to the frequency range $f_{gw} \lesssim 100~\mbox{Hz}$,
with the upper bound corresponding to the peak sensitivity of Einstein
Telescope (ET)~\cite{Sathyaprakash:2012jk} and Cosmic
Explorer~\cite{Evans:2016mbw}. Then  
Eq.~\eqref{frequency} implies
\begin{equation}
\label{glower}
g \lesssim \frac{10^{-8}}{\sqrt{N}} \; ,
\end{equation}
where we set $g_*(T_i)\sim 100$.
One concludes, see Fig.\,\ref{parameters}, that production of detectable GWs is possible in the scenario of DM genesis via the inverse phase transition. The range of values $10^{-9} \lesssim g \lesssim 10^{-8}$ corresponding 
to frequencies $10~\mbox{Hz}\lesssim f_{gw} \lesssim 100~\mbox{Hz}$ will be covered by Einstein Telescope and Cosmic Explorer. DECIGO~\cite{Kawamura:2006up} and BBO~\cite{Corbin:2005ny} will cover 
the frequencies $0.1~\mbox{Hz} \lesssim f_{gw} \lesssim 10~\mbox{Hz}$ and thus probe the range $10^{-11}\lesssim g \lesssim 10^{-9}$. 
For smaller $g$, our model enters the region accessible by LISA~\cite{Audley:2017drz} ($10^{-14} \lesssim g \lesssim 10^{-11}$) and even pulsar timing arrays SKA~\cite{Janssen:2014dka} and IPTA~\cite{IPTA):2013lea} ($g \lesssim 10^{-18}$). Very small $g$ will be also probed by GAIA and THEIA~\cite{Garcia-Bellido:2021zgu}.

In Fig.~\ref{waves} we show the strain of present GWs emitted at
the time $t_i$ for fixed parameters $\beta$ and $N$ and compare it 
with sensitivities of the GW interferometers and pulsar timing arrays. The square of the strain, or the power spectral density $S_h$, is defined from 
\begin{equation}
\Omega_{gw} (t_0) H^2_0 \equiv \frac{2\pi^2 f^3_{gw}}{3} \cdot S_h \; ,
\end{equation}
where $H_0$ is the Hubble constant. As we can see from Fig.~\ref{waves}, future interferometers will be able to probe the model for moderate $\beta/N^2 \lesssim 0.1-10$ corresponding to very small 
$\lambda$.


Emission of GWs is insensitive to the DM mass $M$, since it does not enter the expression for the tension of domain walls, Eq.~(\ref{surface}). 
Nevertheless, in the scenario with the inverse phase transition, the
parameters $g$ and $\beta$ defining the properties of GWs via
Eqs.~(\ref{frequency}) and (\ref{today}) also fix the DM mass, see Eq.~(\ref{constr2}). 
As it follows from Eq.~\eqref{constr2}, where we take $\beta \simeq 1$
and $N \simeq 10$, the potentially observable GWs correspond to
\begin{equation}
\label{Mgw}
M \simeq 5~\mbox{eV} \cdot \left(\frac{g}{10^{-8}} \right)^{7/5} \; ,
\end{equation}
where we put $g_* (T_*) \sim 100$. For example, at LISA peak sensitivity frequency $f_{gw} \simeq 0.01~\mbox{Hz}$, 
the relevant masses are in the axion range $M \simeq 10^{-5}~\mbox{eV}$.

As it follows, the region of parameter space accessible by planned GW detectors involves tiny constants $\lambda$. 
In particular, for $\beta \simeq 1$ and $g \simeq 10^{-8}$, one has $\lambda \simeq 10^{-32}$. Such small values of $\lambda$ are not unnatural insofar as $g$ and $M$ are also small, which is indeed the case according to Eq.~\eqref{Mgw}. In this case, the model is approximately shift symmetric. Assuming that the shift symmetry becomes exact in the limit $g \rightarrow 0$, it follows that
$M \rightarrow 0$ and $\lambda \rightarrow 0$ in the same limit. Here we draw an analogy with the case of axions also enjoying 
an approximate shift symmetry, which guarantees smallness of the axionic mass and self-interaction coupling constant.

{\it Prospects for future.} Multiple different sources may emit GWs in the same frequency range as domain walls in our model. To discriminate between these sources, it is crucial to have information about the spectral shape of GWs produced. This has been obtained in Ref.~\cite{Hiramatsu:2013qaa} by running lattice simulations for the case of constant tension domain walls. Most possibly, it is not completely legitimate to extrapolate these results to our setup 
with the time-dependent tension, and the separate analysis is
required. It is crucial for understanding the smoking gun signature of our DM scenario. 

Finally, let us stress that our estimates rely on some assumptions: i) the field $\chi$ quickly relaxes to its minimum in the broken phase; ii) domain walls enter the scaling regime immediately upon their formation; iii) 
formation of domain walls occurs during radiation-dominated stage. Going beyond these assumptions, one generically 
alters our predictions of GWs properties. We expound this issue in details in future works.

{\it Acknowledgments.} We are indebted to Dra\v zen Glavan for useful discussions. 
The study of gravitational wave signature was supported by the Russian
Science Foundation grant 19-12-00393 (D.~G.). The work of  
E.~B. was supported by the CNRS/RFBR Cooperation program for 2018-2020 n. 1985 ``Modified gravity and black holes: consistent models and experimental signatures''. The work of S.~R. is supported by the Czech Science Foundation GA\v CR, project 20-16531Y. A. V. is supported by the European Regional Development Fund (ESIF/ERDF) and the Czech Ministry of Education, Youth and Sports (M\v SMT) through the Project CoGraDS- CZ.02.1.01/0.0/0.0/15 003/0000437.

\end{document}